\documentstyle[12pt]{article}
\begin{document}
\begin{titlepage}
\begin{flushright}
TU- 567\\
cond-mat/9906377\\
To be published in Phys. Lett. A
\end{flushright}
\ \\
\begin{center}
\LARGE
{\bf
 Composability \\
   and  \\
Generalized Entropy
}
\end{center}
\ \\
\begin{center}
\Large{
M.Hotta${}^\ast$ and I.Joichi${}^\dagger$    }\\
{\it
${}^\ast$
Department of Physics,Tohoku University,\\
Sendai 980-8578,Japan\\
hotta@tuhep.phys.tohoku.ac.jp \\

\ \\

${}^\dagger$
School of Science and Engineering,\\
 Teikyo University,\\
Toyosatodai 1-1, Utsunomiya 320-8551, Japan \\
joichi@umb.teikyo-u.ac.jp
}
\end{center}

\begin{abstract}
We address in this paper how tightly
the composability nature of systems:
$$
S_{A+B}
=\Omega \left(S_A ,S_B \right)
$$
constrains
 definition of generalized entropies  and investigate
 explicitly
 the composability in some ansatz of the entropy form.
\end{abstract}

\end{titlepage}

\section{Introduction}

\ \\

For the recent decade generalization of Boltzmann-Gibbs
 statistics has  gradually  attracted attention
 and it has already been recognized that
 many interesting phenomena in various fields really need
  modification of their entropies for their precise descriptions
 \cite{hp,rv}.

Even though the entropy is modified,
 it is expected to possess several fundamental properties
 in order to express macroscopic aspects of physical systems.
 Many efforts to investigate the properties, including
 the concavity and the H-theorem,
 have been made so far and shed light on
 understanding of
the generalized  thermodynamics itself \cite{hp,m,r}.

We shall analyze  in this paper
the composability \cite{c,rv}
 which is a candidate of natural property
 of the generalized entropy.
Let us consider two systems A and B of the same material structure
 with  different numbers of states.
 Their entropies are denoted by
\begin{eqnarray}
&&
S_A =S_A (P^A_i ),
\\
&&
S_B =S_B (P^B_j ).
\end{eqnarray}
We can always regard the two systems as one composite system
 $A+B$ .
So the total entropy of the system $A+B$ is  requested
 to exist with some form:
\begin{eqnarray}
S_{A+B} =S_{A+B} (P_{ij} ),
\end{eqnarray}
where $P_{ij}$ is probability related to the composite system and
 the index \\
$i$( $j$ ) corresponds to
 degree of freedom of the system $A$($B$).
Now let us keep focusing on  cases in which
 the interaction between the systems $A$ and  $B$
 is so small that it can be neglected.
Due to this the probability of the composite system
may take the value as
\begin{eqnarray}
P_{ij} = P^A_i P^B_j .
\end{eqnarray}
Then the definition of the composability
can be given as following.
\begin{eqnarray}
S_{A+B} ( P^A_i P^B_j )
=\Omega \left(S_A ,S_B \right),
\end{eqnarray}
where $\Omega$ is a function
with two arguments $S_A$ and $S_B$.
The function $\Omega$ should not depend on
 the number of states of the systems.
This definition means
that the total entropy is  deterministic macroscopically
and can be built of just
the two macroscopic quantities, $S_A$ and $S_B$.

We point out that
 the composability constrains
the form of the generalized entropy quite tightly.
In Section 2, we discuss
an ansatz:
$$
S (P_i )
=C+\sum_{i} \phi(P_i )
$$
 as a modification
 of the entropy definition and
 force the ansatz to respect the composability.
 Consequently uniqueness
 of the Tsallis entropy, up to constant and factor, is proven 
 under some assumptions.

In Section 3, we extend
the ansatz of the entropy into more complicated one
and
find a species of the generalized entropy holding the composability.

\section{Composability and Tsallis Entropy}

\ \\

It is a quite interesting problem to clarify how strongly
the composability constrains the form of the generalized
entropy definition itself.
 In this section,
we concentrate our attention
on an ansatz of the generalized entropy definition
 as follows.
\begin{eqnarray}
S (P_i )
=C+\sum_{i} \phi(P_i ),
\label{000}
\end{eqnarray}
where $C$ is a constant,
$\phi$ is an unfixed smooth
 function independent of the number of states
of the system,
$P_i$ is probability corresponding to Event $i$.
 Taking the complete sum of all the probabilities,
\begin{eqnarray}
\sum_i P_i =1
\end{eqnarray}
should  hold by definition.

Here it should be noted that the function $\phi$ satisfies
\begin{eqnarray}
\phi(0)=0.\label{1}
\end{eqnarray}
To prove this, let us consider a system with $N+1$
 states.
By taking $P_{N+1} =0$,
 the system is straightforwardly
reduced into a system with $N$ states.
Then the entropies  $S_{N+1} (P_{N+1}=0)$ and $S_{N}$
naturally coincide with each other:
$$
S_{N+1} (P_{N+1}=0)=S_{N}.
$$
{F}rom this  eqn (\ref{1}) is found easily.

Due to the definition (\ref{000})
 the entropies of each systems are written down explicitly
as follows.
\begin{eqnarray}
&&
S_{A+B} (P_{ij} )
=C+\sum_{i=1}^N \sum_{j=1}^M \phi(P_{ij} ),\label{2}
\\
&&
S_{A} (P^A_i )
=C+\sum_{i=1}^N \phi(P^A_i ),
\\
&&
S_{B} (P^B_j )
=C+\sum_{j=1}^M \phi(P^B_j ),
\end{eqnarray}
where $N$($M$) is the number of states of the system $A$($B$).

Now we address a problem what kind of constraints
the composability imposes on the functions $\Omega$ and $\phi$.

Firstly let us analyze the case with $N=2$ and $M=2$ .
The probabilities for the system $A$($B$) are denoted by
$a$ and $1-a$ ($b$ and $1-b$) where $a$($b$) is a parameter
 with $0\leq a \leq 1$ ($0\leq b\leq 1$).
Then the following expression for the combined entropy
 is obtained from the definition (\ref{2}).
\begin{eqnarray}
&&
\Omega (S_A ,S_B )
\nonumber\\
&&
=
C+
\phi(ab)
+\phi(a(1-b))
+\phi((1-a)b)
+\phi((1-a)(1-b)).
\label{3}
\end{eqnarray}
Also the entropies for $A$ and $B$ read as follows.
\begin{eqnarray}
&&
S_A =C+\phi(a) +\phi(1-a),
\label{4}\\
&&
S_B =C+\phi(b) +\phi(1-b).
\label{5}
\end{eqnarray}
It is noticed here that
 the set of eqns (\ref{3}),(\ref{4}) and (\ref{5})
 is just a parametric representation of
 the function $\Omega (S_A, S_B)$
 by two free parameters $a$ and $b$.
Thus this fact tells us that
the function $\Omega$ is determined by giving the function $\phi$,
 as naturally expected.
Therefore we do not care about $\Omega$-hunt any more and
can devote ourselves  later to search for the function $\phi$.

Next let us consider
the case with $(N,M)=(N, 2)$  and  the case with $(N,M)=(N,3)$
simultaneously. For the system A with N states,
 let us make the probabilities  denoted by $P_i$ with
$$
\sum^N_{i=1} P_i =1.
$$
And for the system B with 2 states,
let $x$ and $1-x$ denote the probabilities.
Then the entropies are expressed using these variables
as follows.
\begin{eqnarray}
&&
S_A =C+\sum^N_{i=1} \phi(P_i ),
\\
&&
S_B =C+\phi(x) +\phi(1-x).
\end{eqnarray}
Meanwhile for another remaining system $B'$ with 3 states
let us make $y$ ,$z$ and $1-y-z$  denote the probabilities.
Then the entropy  is written as
\begin{eqnarray}
S_{B'} =C+\phi(y) +\phi(z)+\phi(1-y-z).
\end{eqnarray}
We try to adjust the two entropies $S_B$ and $S_B '$ to
 take the same value:
\begin{eqnarray}
S_B =S_{B'}.\label{6}
\end{eqnarray}
 It is always possible
to realize the situation by substituting a suitable function
\begin{eqnarray}
x=x(y,z)
\end{eqnarray}
into the entropy $S_B$. 
Then the condition (\ref{6}) is reexpressed as
\begin{eqnarray}
\phi\left( x(y,z)\right) +\phi \left( 1-x(y,z)\right)
=
\phi(y) +\phi(z) +\phi (1-y-z).\label{7}
\end{eqnarray}
Eqn (\ref{7}) makes the function $x(y,z)$ fixed
 after determination of the function $\phi(x)$ .
Due to adoption of the above function $x(y,z)$  
the relation:
\begin{eqnarray}
\Omega (S_A ,S_B  )
=\Omega(S_A ,S_{B'} )
\end{eqnarray}
 trivially holds and is rewritten explicitly as follows.
\begin{eqnarray}
\sum^N_{i=1} \phi(xP_i ) +\sum^N_{i=1} \phi( (1-x)P_i )
=
\sum^N_{i=1} \phi(yP_i )
+\sum^N_{i=1} \phi(zP_i )
+\sum^N_{i=1} \phi ((1-y-z)P_i ).\label{8}
\end{eqnarray}
Eqn (\ref{8}) must be satisfied for arbitrary
 probability variables, $y$,$z$ and $P_i$.
Therefore we are able to  set
$$
P_i =\frac{1}{N}
$$
and
$$y=z.$$
Then a compact version of the master equation (\ref{8})
is induced as follows.
\begin{eqnarray}
&&\phi\left( \frac{x}{N}\right)
+\phi \left(\frac{1-x}{N} \right)\nonumber\\
&&=
2\phi\left(\frac{y}{N}\right)
+\phi \left(\frac{1-2y}{N}\right),
\label{0}
\end{eqnarray}
for $N=2,3,\cdots$.
It is noticed that
eqn (\ref{7}) with $y=z$ corresponds to the $N=1$ case
of eqn (\ref{0}).
Thus in later discussion we shall deal with
  the equation
 as a special case of eqn (\ref{0}) for positive integer $N$. 
Just for convenience we use later not $x=x(y,y)$ but $y=y(x)$,
the inverse function of $x=x(y,y)$ .

Next we shall prove that
 the index $N$ in eqn (\ref{0}) can be extended into
 a continuous variable.
For this purpose let us take the following distribution
 for $P_i$ of the system $A$.
\begin{eqnarray}
&&
P_1 = p,
\\
&&
P_i =\frac{1-p}{N-1}\ \ (i=2,\cdots, N).
\end{eqnarray}
Substituting this into eqn (\ref{8})
 yields a new relation:
\begin{eqnarray}
&&
\phi\left( xp \right)
+
\phi\left( (1-x) p\right)
\nonumber\\
&&
+(N-1)\phi \left(x\frac{1-p}{N-1} \right)
+(N-1)\phi \left((1-x)\frac{1-p}{N-1} \right)
\nonumber\\
&&
=
2\phi\left( yp \right)
+
\phi\left( (1-2y) p\right)
\nonumber\\
&&
+2(N-1)\phi \left(y\frac{1-p}{N-1} \right)
+(N-1)\phi \left((1-2y)\frac{1-p}{N-1} \right).
\label{9}
\end{eqnarray}
By differentiating with respect to $p$ iteratively in eqn (\ref{9})
and taking $p=1/N$,
the following relations are obtained.
\begin{eqnarray}
&&
x^n \phi^{(n)} \left( \frac{x}{N}\right)
+(1-x)^n \phi^{(n)} \left(\frac{1-x}{N} \right)
\nonumber\\
&&
=
2y^n \phi^{(n)} \left(\frac{y}{N}\right)
+(1-2y)^n \phi^{(n)} \left(\frac{1-2y}{N}\right),
\label{10}
\end{eqnarray}
where $n=0,2,3,4,\cdots$ and
$\phi^{(n)}(x)$ is the $n$-th derivative of the function
$\phi(x)$ with respect to $x$.
Unfortunately the relation (\ref{10})
corresponding to $n=1$ does not appear
due to a coincidence that the derived equation for $n=1$ becomes
 trivial when $p=1/N$ is taken.
However we can remedy the lack using another source
provided that $\phi'(0)$ exists.
Let us  replace $N$ with $N+1$ in eqn (\ref{9})
and substituting
\begin{eqnarray}
&&
P_i =\frac{1-p}{N},\ \ (i=1,2,\cdots ,N)
\\
&&
P_{N+1} =p,
\end{eqnarray}
 into the equation.
 It turns out that the following equation holds.
\begin{eqnarray}
&&
\phi\left( xp \right)
+
\phi\left( (1-x) p\right)
\nonumber\\
&&
+N\phi \left(x\frac{1-p}{N} \right)
+N\phi \left((1-x)\frac{1-p}{N} \right)
\nonumber\\
&&
=
2\phi\left( yp \right)
+
\phi\left( (1-2y) p\right)
\nonumber\\
&&
+2N\phi \left(y\frac{1-p}{N} \right)
+N\phi \left((1-2y)\frac{1-p}{N} \right).
\label{11}
\end{eqnarray}
Assuming the existence of $\phi' (0)$,
we can set $p=0$
after the differentiation with respect to $p$ in eqn (\ref{11}).
 After some manipulation
 the missing equation (\ref{10}) for $n=1$ is really derived:
\begin{eqnarray}
&&
x\phi' \left( \frac{x}{N}\right)
+(1-x) \phi' \left(\frac{1-x}{N} \right)
\nonumber\\
&&
=
2y \phi' \left(\frac{y}{N}\right)
+(1-2y) \phi' \left(\frac{1-2y}{N}\right) .
\nonumber
\end{eqnarray}
After all it  have been proven  that
\begin{eqnarray}
A_n &=& x^n \phi^{(n)} \left( \frac{x}{N}\right)
+(1-x)^n \phi^{(n)} \left(\frac{1-x}{N} \right)
\nonumber\\
&&
-
2y^n \phi^{(n)} \left(\frac{y}{N}\right)
-(1-2y)^n \phi^{(n)} \left(\frac{1-2y}{N}\right)
\nonumber\\
&=&0,
\label{12}
\end{eqnarray}
where $n$ takes all non-negative integer values.
Next let us introduce a deviation parameter $\Delta N$ which takes
 real values and add $\Delta N$ to $N$ .
Then we  can show that a function defined as
\begin{eqnarray}
\Psi(N+\Delta N)&=&
\phi\left( \frac{x}{N+\Delta N}\right)
+\phi \left(\frac{1-x}{N+ \Delta N} \right)
\nonumber\\
&&
-
2\phi\left(\frac{y}{N+\Delta N}\right)
-\phi \left(\frac{1-2y}{N +\Delta N}\right)
\label{250}
\end{eqnarray}
 vanishes for arbitrary values of
the continuous parameter  $\Delta N$ as follows.
The function $\Psi$ may be Taylor-expanded as
\begin{eqnarray}
\Psi(N+\Delta N)=
\sum_{n=0}^\infty \frac{1}{n!}\Psi_n (N) \Delta N^n .
\end{eqnarray}
After some manipulation it is shown  that
all the $\Psi_n $ are equal to linear combinations of $A_n$
 like that
\begin{eqnarray}
&&
\Psi_0 =A_0,
\\
&&
\Psi_1 =-\frac{1}{N^2} A_1,
\\
&&
\Psi_2 =\frac{2}{N^3}A_1 +\frac{1}{N^4} A_2 ,
 \\
&&
\Psi_n  =\sum^n_{k=1} C^{(n)}_k (N) A_k .\ \ ( n \geq 3)
\end{eqnarray}
Thus from eqn (\ref{12})
\begin{eqnarray}
\Psi_n  =0
\end{eqnarray}
must be satisfied for $n=0,1,2,\cdots$ and $\Psi(N+\Delta N)$
 vanishes. 
This means that eqn (\ref{0}) must hold for arbitrary
 real number $N$.
By virtue of this fact,
 our task has now been rather simplified and
 is  to find the two functions $\phi(x)$ and $y(x)$
 that satisfy
 eqn (\ref{0}) for arbitrary real parameter, $N$ .

{F}rom eqn (\ref{0}),
we are able to prove under some conditions that
 the generalized entropy (\ref{000})
 possessing the composability is equal to the Tsallis entropy,
 up to constant and factor.
Actually
we give two independent proofs  in the following.

Firstly we assume a rather general ansatz for
the function $\phi(x)$ as follows.
\begin{eqnarray}
\phi(x) =\sum_{\nu=\nu_{min} }
\sum^\infty_{n=0} \phi_n (\nu )x^{\nu +n},
\label{13}
\end{eqnarray}
where $\nu$ are real numbers and
$\nu_1 -\nu_2 \neq 0\ (mod\ integer)$ if $\nu_1$ and $\nu_2$
 appear in the sum. Also
 the coefficient $\phi_0 (\nu )$ does not vanish.
 In eqn (\ref{13}), there exists the minimum positive value for
 $\nu$;$\nu_{min} \geq 1$
because we must ensure that $\phi(0)=0$
 and the existence of $\phi'(0)$.
 Substituting eqn (\ref{13}) into eqn (\ref{0}) and
taking the expansion on $1/N$
 for large $N$ yield
\begin{eqnarray}
\sum_{\nu=\nu_{min} } \sum^\infty_{n=0}
\left(\frac{1}{N}\right)^{\nu +n}
\phi_n (\nu )
\left[
x^{\nu +n}
+(1-x)^{\nu +n}
-2y^{\nu +n}
-(1-2y)^{\nu +n}
\right]
=0.
\end{eqnarray}
By comparing each of the coefficients in the expansion on $1/N$,
it is straightforwardly obtained that
\begin{eqnarray}
\phi_n (\nu )
\left[
x^{\nu +n}
+(1-x)^{\nu +n}
-2y^{\nu +n}
-(1-2y)^{\nu +n}
\right]
=0.
\end{eqnarray}
If we have two or more non-zero coefficients
$\phi_{n_i} (\nu_i )$, the function $y=y(x)$ must satisfy
 the following equations for all the $i$'s.
\begin{eqnarray}
x^{\nu_{i}+n_i } +(1-x)^{\nu_{i} +n_i }
=2y^{\nu_{i}  +n_i } +(1-2y)^{\nu_{i}+n_i }.
\label{600}
\end{eqnarray}
However, this is clearly impossible and we fails to find  
 any solution $y=y(x)$ like that. Thus we conclude that 
 there exists only one
non-vanishing coefficient $\phi_0 (\nu_{min} )\neq 0$.
Consequently from eqn (\ref{13})
the function $\phi(x)$ is uniquely given by
\begin{eqnarray}
\phi(x) =B x^{q }
\label{15}
\end{eqnarray}
where $B=\phi_0 (q)$ is some constant, $q=\nu_{min} \geq 1$
and
the function $y=y(x)$ is determined by
the following equation;
\begin{eqnarray}
x^{q } +(1-x)^{q }
=2y^{q } +(1-2y)^{q }
\end{eqnarray}
due to eqn (\ref{600}).
Substituting (\ref{15}) into
 definition of the generalized entropy (\ref{000})
and introducing constants $S_o$ and $D$ like
\begin{eqnarray}
&&
C=S_o -\frac{D}{1-q},
\\
&&
B=\frac{D}{1-q},
\end{eqnarray}
we get the main result of this section;
\begin{eqnarray}
S (P_i )
=S_o +D S_{q}
\label{17}
\end{eqnarray}
where
\begin{eqnarray}
S_{q}=
-\frac{1-\sum_{i} P_i^q }{1-q}
\label{300}
\end{eqnarray}
 is just the  Tsallis entropy \cite{t} and
if we are able to take  $q\rightarrow 1$
beyond the assumption that $\phi'(0)$ exists,
Boltzmann-Gibbs entropy:
\begin{eqnarray}
S_{q=1}=-\sum_i P_i \ln P_i
\end{eqnarray}
 is reproduced.
One can check easily that the entropy (\ref{17}) actually  has
 the composability and its additive law of the entropy
 is modified as
\begin{eqnarray}
S_{A+B}
=S_A +S_B -S_o+\frac{1-q}{D}(S_A -S_o )(S_B -S_o ) .
\end{eqnarray}
We expect that
even if the ansatz (\ref{13}) for the function $\phi$ is extended
to a more complicated form,
eqn(\ref{0}) will prevent the result derived above from changing.

Besides the above proof
we can show another one in order to argue the uniqueness.
Let us introduce
two independent variables $X$ and $Y$ as follows.
\begin{eqnarray}
&&
X=\phi(x) +\phi(1-x) -2\phi\left( \frac{1}{2} \right),
\\
&&
Y=2\phi(y) +\phi(1-2y) -2\phi\left( \frac{1}{2} \right),
\end{eqnarray}
where
$x=1/2$ ($y=1/2$) corresponds to
$X=0$ ($Y=0$).
Let us consider again the function $\Psi$, eqn (\ref{250}).
The function depends on only three independent
 variables, ($N$, $X$ , $Y$) or ($N$,$x$,$y$).
 Let us assume later that
the function $\Psi$ may be Maclaurin-expanded
in terms of $X$ and $Y$ as
\begin{eqnarray}
\Psi=\sum_{k=0}^\infty \sum_{l=0}^\infty
 \frac{1}{k! l!} \tilde{\Psi}_{kl} (N) X^k Y^l .
\label{261}
\end{eqnarray}
It is worthwhile to note a simple fact that the function $\Psi$
 is  obtained by adding a function of $x$:
$$
\phi\left( \frac{x}{N}\right)
+\phi \left(\frac{1-x}{N} \right)
$$
 and a function of $y$:
$$
-
2\phi\left(\frac{y}{N}\right)
-\phi \left(\frac{1-2y}{N }\right),
$$
that is, takes the separate-variables form with respect to
 the variables $x$ and $y$.
Moreover eqn (\ref{0}) must hold when
$$
\phi(x)+\phi(1-x)-2\phi(y)-\phi(1-2y)
=X(x)-Y(y)=0
$$
is satisfied.  Because of  these two conditions
the coefficients $\tilde{\Psi}_{kl}$ are drastically constrained
 and eqn (\ref{261}) can be replaced into the following equation.
\begin{eqnarray}
&&
\phi\left( \frac{x}{N}\right)
+\phi \left(\frac{1-x}{N} \right)
-
2\phi\left(\frac{y}{N}\right)
-\phi \left(\frac{1-2y}{N }\right)
\nonumber\\
&&
=\sum^\infty_{k=0} \tilde{\Psi}_k (N) (X^k -Y^k ) \label{251},
\end{eqnarray}
where $\tilde{\Psi}_k$ are
 members of the coefficients $\tilde{\Psi}_{kl}$ that survive
 even after imposing the above two conditions.
Therefore we acquire the two independent relations as
\begin{eqnarray}
&&
\phi\left( \frac{x}{N}\right)
+\phi \left(\frac{1-x}{N} \right)
=
\sum^\infty_{k=0} \tilde{\Psi}_k (N)X^k
 + \delta (N) ,
\label{28}
\\
&&
2\phi\left(\frac{y}{N}\right)
+\phi \left(\frac{1-2y}{N }\right)
=
\sum^\infty_{k=0} \tilde{\Psi}_k (N) Y^k
 +\delta (N) \label{29},
\end{eqnarray}
where $\delta(N)$ is a undetermined function of $N$
  needed when we separate the variables $x$ and $y$
in eqn (\ref{251}).
 The function $\delta$ can be always absorbed
by the coefficient $\tilde{\Psi}_0$,
 so  we  fix  $\delta =0$ later.
The coefficients $\tilde{\Psi}_k$ are  calculated by
 repeatedly differentiating
the left-hand-side of eqn (\ref{28}) on $X$
 and setting $X=0$, that is, $x=1/2$ as follows.
\begin{eqnarray}
\tilde{\Psi}_0 &=&2\phi\left( \frac{1}{2N} \right),
\label{30}
\\
\tilde{\Psi}_1 &=&\frac{\phi^{(2)} \left(\frac{1}{2N}\right) }
        {N^2  \phi^{(2)} \left(\frac{1}{2}\right) }.
\label{31}
\end{eqnarray}
 On the other hand, the same coefficients must be obtained
  from repeatedly
differentiation of eqn (\ref{29}) with respect to $Y$ and setting
 $Y=0$, or $y=1/2$.
In fact substituting $y=1/2$ into eqn (\ref{29})
 reproduces eqn (\ref{30}).
 Meanwhile a non-trivial constraint is generated for
the $\tilde{\Psi}_1$ calculation.
 The form of $\tilde{\Psi}_1$ obtained this time
 is as follows.
\begin{eqnarray}
\tilde{\Psi}_1 &=&
\frac{1}{N}
\frac{
\phi'\left(\frac{1}{2N} \right)
-\phi'\left(0 \right)
}{
\phi'\left(\frac{1}{2} \right)
-\phi'\left(0\right)
}
.\label{331}
\end{eqnarray}
Equating eqn (\ref{31}) and eqn (\ref{331}),
and replacing $1/(2N)$ into $x$, it can be proven  that
 the function $\phi(x)$ must satisfy the following
 differential equation.
\begin{eqnarray}
x^2 \frac{d^2 \phi}{dx^2}
-\frac{1}{2}
\frac{\phi^{(2)}\left(\frac{1}{2} \right)}
{\phi '\left(\frac{1}{2} \right)
-\phi '\left(0\right) }
 x \frac{d \phi}{dx}
=  -\frac{1}{2}
\frac{\phi'\left(0\right)
\phi^{(2)} \left(\frac{1}{2} \right)}
{\phi '\left(\frac{1}{2} \right)
-\phi '\left(0\right) } x .
\end{eqnarray}
By integrating the equation,
 it is shown that the function $\phi(x)$  take the form of
\begin{eqnarray}
\phi(x) =\phi'(0) x + C_1 +C_2 x^q ,
\label{34}
\end{eqnarray}
where $C_1$ and $C_2$ are integrated constants and
$$
q=1+\frac{1}{2}
\frac{\phi^{(2)}\left(\frac{1}{2} \right)}
{\phi '\left(\frac{1}{2} \right)
-\phi '\left(0\right) }.
$$
Substituting eqn (\ref{34}) into eqn (\ref{0}) ,
  the same result of the first analysis is reproduced:
\begin{eqnarray}
\phi(x)\propto x^q .\ \ \ (q\geq 1)
\end{eqnarray}

Consequently it  has been proven,
 assuming  the ansatz of  the function
 $\phi$ (\ref{13}) or of the function $\Psi$ (\ref{261})
 besides the existence of $\phi'(0)$,
 that the composability requests
 the Tsallis entropy uniquely,
up to the constant $S_o$ and  the factor $D$,
 in eqn (\ref{000}).
This means that the composability can play
 a quite significant role in
the generalized entropy search.

We  comment that
 dos Santos also proved
 uniqueness of the Tsallis entropy,
assuming the Tsallis pseudo-additivity:
\begin{eqnarray}
S_{A+B} =S_A +S_B +(1-q)S_A S_B
\label{500}
\end{eqnarray}
and other several conditions \cite{d}.
 It should be stressed here that
the pseudo-additivity (\ref{500}) is not assumed a priori
 in our analysis.

\section{Extension}

\ \\

In this section we try to extend the analysis
in Section 2 for more complex definition of
 the generalized entropies.
 Instead of the ansatz in eqn (\ref{000}),
we adopt here the following  form.
\begin{eqnarray}
S =S\left(
\sum_i P^{q_1}_i ,\sum_i P^{q_2}_i ,\cdots,\sum_i P^{q_K}_i
\right),
\label{26}
\end{eqnarray}
where $S$ is given as a Laurent series:
\begin{eqnarray}
S(X_1 ,X_2 ,\cdots, X_K )
=\sum^\infty_{i_1 =-\infty}
\sum^\infty_{i_2 =-\infty}\cdots\sum^\infty_{i_K =-\infty}
S_{i_1 , i_2 ,\cdots ,i_K }
X^{i_1 }_1 X^{i_2}_2 \cdots X^{i_K}_K \label{400}.
\end{eqnarray}
 Here the power exponents $q_m (\neq 1)$
take different values each other and
 the coefficients $S_{i_1 ,i_2 ,\cdots ,i_K }$ are
independent of the number of states
 of the systems.
Also $K$ is assumed finite positive integer.

We can repeat straightforwardly the analysis of the composability
using the cases with $(N,2)$ and $(N,3)$ in Section 2.
Probabilities for the system $A$ with $N$ states,
$B$ with 2 states and $B'$ with 3 states are denoted
 in the same way of Section 2.
 Then it can be pointed out that
the composability calls for the
 existence  of a function $y=y(x)$ which satisfies that
\begin{eqnarray}
&&S(f_1 (x) Q_1  , f_2 (x) Q_2  ,\cdots, f_K (x) Q_K  )
\nonumber\\
&&
=
S(g_1 (y) Q_1  , g_2 (y) Q_2  ,\cdots, g_K (y) Q_K  ),
\label{20}
\end{eqnarray}
where
\begin{eqnarray}
&&
Q_k =\sum^N_{i=1} P^{q_k}_i ,\\
&&
f_k (x) = x^{q_k} +(1-x)^{q_k}, \\
&&
g_k (y) =2y^{q_k} +(1-2y)^{q_k}.
\end{eqnarray}
If $N\geq K$ is taken,
 it is noticed  from  counting the degree of freedom
 that all the $Q_k $ are independent parameters each other.
This fact may sound trivial but shows us a significant result.
Using eqn (\ref{400}), the relation (\ref{20}) is rewritten as
\begin{eqnarray}
\sum
S_{i_1 , i_2 ,\cdots ,i_K }
\left[\prod^K_{k=1}f_k (x)^{i_k}
-\prod^K_{k=1} g_k (y)^{i_k} \right]
Q_1^{i_1 } Q_2^{i_2} \cdots Q_K^{i_K}
=0 .
\end{eqnarray}
Thus
\begin{eqnarray}
S_{i_1 , i_2 ,\cdots ,i_K }
\left[\prod^K_{k=1}f_k (x)^{i_k}
-\prod^K_{k=1} g_k (y)^{i_k} \right]  =0 \label{404}
\end{eqnarray}
 must hold simultaneously for all the indices.
In order not to overfix the function $y(x)$
 due to two or more constraints in eqn (\ref{404}),
the coefficients $S_{i_1 , i_2 ,\cdots ,i_K }$
 vanish except
\begin{eqnarray}
S_{jl_1  ,jl_2 ,\cdots ,jl_K }
\ \ \ \ (j:integer )
\end{eqnarray}
where $l_k \ (k=1\sim K)$ are
integers chosen arbitrarily. Then the function $y$
 is defined by the following equation.
\begin{eqnarray}
\prod^K_{k=1}f_k (x)^{l_k}
=\prod^K_{k=1} g_k (y(x))^{l_k} .
\end{eqnarray}
 Introducing a function $F(X)$ as
\begin{eqnarray}
F(X)=\sum^\infty_{j=-\infty} S_{jl_1  ,jl_2 ,\cdots ,jl_K } X^j ,
\end{eqnarray}
the  entropy now reads
\begin{eqnarray}
S=F\left(
\prod^K_{k=1}
\left(\sum_i P^{q_k}_i \right)^{l_k}
\right).\label{405}
\end{eqnarray}

It can be checked straightforwardly using
\begin{eqnarray}
\sum_{ij} \left(P^A_i P^B_j \right)^q
=\left( \sum_i \left( P^A_i \right)^q \right)
\left(\sum_j \left( P^B_j  \right)^q \right)
\label{25}
\end{eqnarray}
that the result (\ref{405})
really possesses the composability.
 Actually a parametric expression of the composability
 is explicitly given as
\begin{eqnarray}
&&
S_{A+B} =F\left( ab \right) ,
\\
&&
S_A = F \left(a \right),
\\
&&
S_B= F \left(b \right),
\end{eqnarray}
where
\begin{eqnarray}
a=\prod^K_{k=1}
\left(\sum_i (P_{Ai})^{q_k} \right)^{l_k}
\end{eqnarray}
and
\begin{eqnarray}
b=
\prod^K_{k=1}
\left(\sum_j (P_{Bj})^{q_k} \right)^{l_k}
\end{eqnarray}
are its free parameters.

Note that the form of eqn (\ref{405}) involves
  a modified Tsallis entropy:
\begin{eqnarray}
\tilde{S}_q =-\frac{1}{1-q}
\frac{1-\sum_i P_i^q }{\sum_i P^q_i}
\end{eqnarray}
 which was proposed by Rajagopal and Abe \cite{r-a}
and is known to be equipped with the composability nature.

In conclusion, it has been clarified that
if one imposes  the composability
 on the generalized entropy $S$ (\ref{26}),
 the form of the entropy is restricted to eqn (\ref{405}).

Finally we would like to comment that
 the composablity we argue here is just a conjecture so far
and it may be possible to propose attractive 
 non-composable entropies which express significant aspects
 of some exotic systems.  Actually Anteneodo and Plastino
\cite{ap} give quite an interesting entropy form which
 possesses a lot of plausible physical properties
 but is clearly non-composable.

\ \\
\ \\
\ \\
{\bf Acknowledgement}\\

 We would like to thank S. Abe, M. Morikawa, O. Iguchi and
 K. Sasaki for fruitful discussions and informing us of
 related references. I.J. whould also like to thank 
T. Yokobori and A. T. Yokobori, Jr for hospitality.

\end{document}